\documentclass[twocolumn,floatfix,prb,aps,showpacs]{revtex4}
\usepackage{graphicx,amsmath,amssymb}

\begin{document}

\title{The effect of the electron-electron interaction on the Lifshitz transition density in bilayer graphene}
\author{Csaba T\H oke and Vladimir I. Fal'ko}
\affiliation{Department of Physics, Lancaster University, Lancaster, LA1 4YB, United Kingdom}
\date{\today}

\begin{abstract}
We study the renormalization of the effective mass and trigonal warping of bilayer graphene by the electron-electron interaction.
One consequence of such a renormalization in the low-energy bands of a bilayer crystal consists of a small reduction of
the critical density of the Lifshitz transition (the crossover between the single-pocket and four-pocket topology of the Fermi surface).
\end{abstract}


\maketitle

\newcommand{\bsigma}{\mbox{\boldmath$\sigma$}}

\begin{figure}[htbp]
\begin{center}
\includegraphics[width=0.8\columnwidth, keepaspectratio]{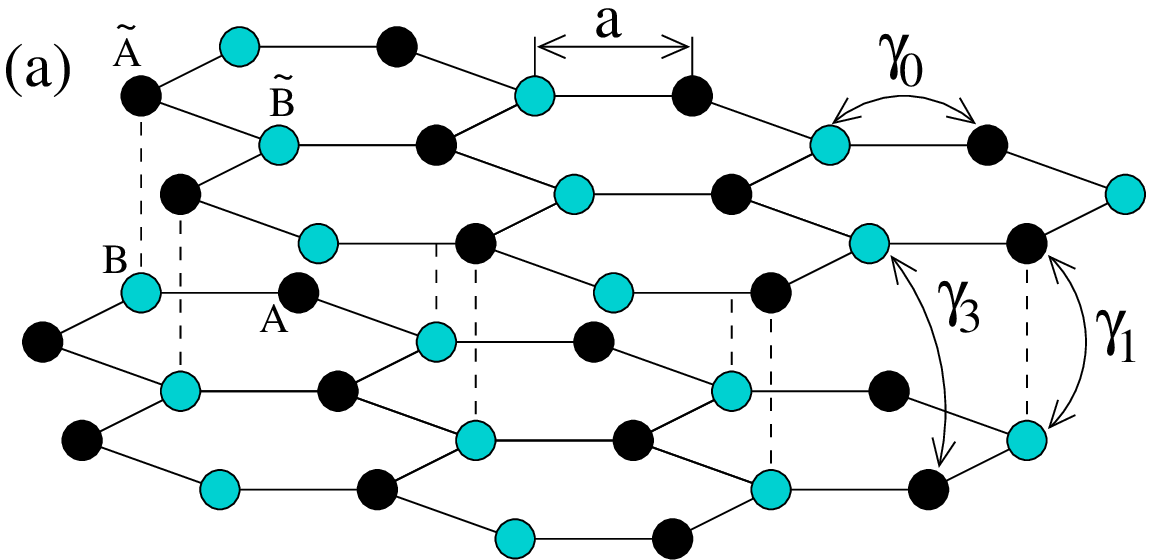}

\includegraphics[width=0.8\columnwidth, keepaspectratio]{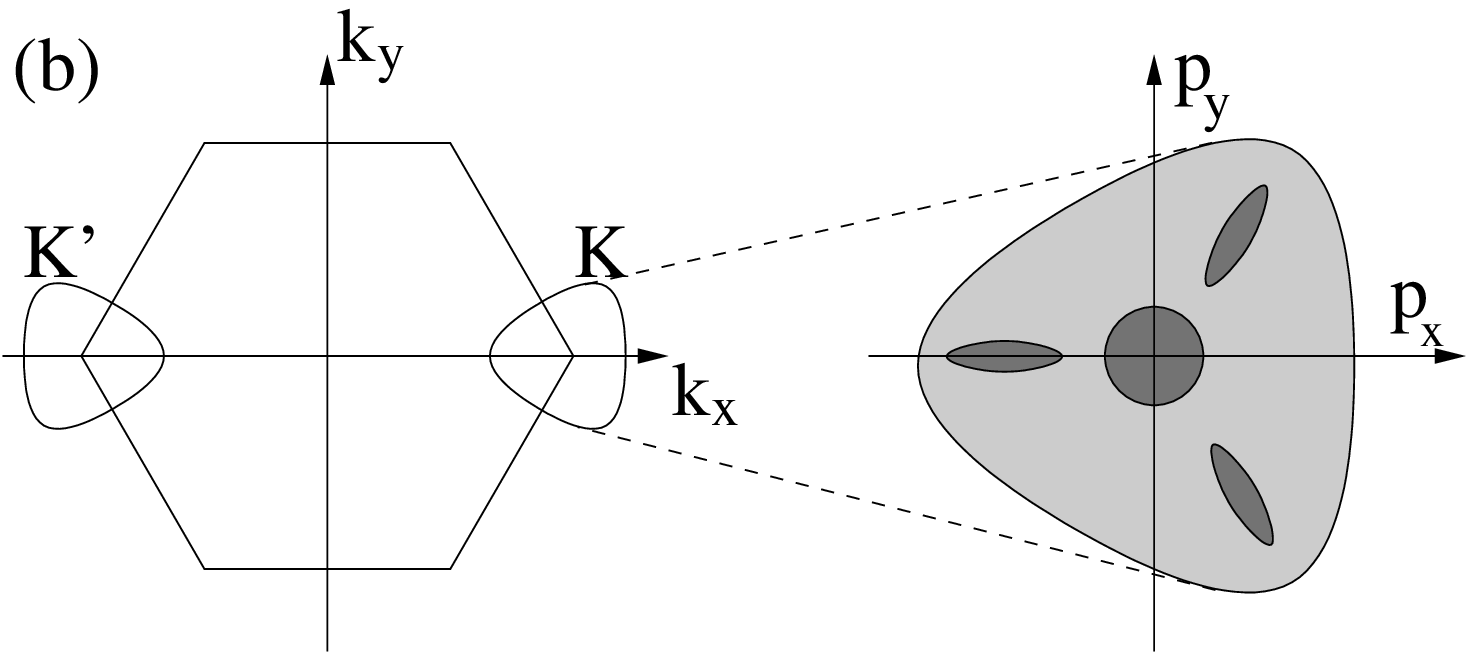}
\end{center}
\caption{\label{intro}
(Color online)
(a) The crystalline structure of bilayer graphene.
(b) The first Brillouin zone and the constant energy cuts about the $K$, $K'$ points.
The dark shaded area represents the four pockets in momentum space filled by electrons in the gas with $|n|<n_L$ below the Lifshitz transition density,
the light shaded area is the simply connected Fermi line for higher densities.
}
\end{figure}

The bilayer graphene crystal is one of several allotropic forms of carbon one can fabricate using micro-mechanical cleavage of bulk graphite\cite{Novoselov}.
The crystalline structure of bilayer graphene is derived from the Bernal stacking\cite{Bernal} of layers in the ``mother'' graphite crystal.
It is shown in Fig.~\ref{intro}(a) as two hexagonal carbon lattices with inequivalent sites $A,B$
(bottom layer) and $\widetilde A,\widetilde B$ (top layer) such that the $\widetilde A$ sites are above the $B$ sites,
while $A$ ($\widetilde B$) is located above (below) the midpoint of a hexagon on the other layer.
Theoretical studies\cite{McCann}, transport\cite{Novoselov}, ARPES\cite{Ohta} and optical\cite{Kuz} characterization of bilayer graphene have
indicated that this material is a gapless
semiconductor with two split-bands separated by $\pm\gamma_1$, where $\gamma_1$ is the closest-neighbor interlayer ($\widetilde AB$) coupling, and
two ``low-energy'' almost parabolic bands with effective mass
determined by both $\gamma_1$ and the intralayer coupling $\gamma_0$.

Detailed tight-binding model studies\cite{McCann} of bilayer graphene taking into account the next-neighbor ($\widetilde BA$) interlayer coupling $\gamma_3$ have
indicated that the dispersion of the low-energy bands in it is strongly anisotropic at small momenta $p\sim\gamma_3\gamma_1/(\gamma_0^2a)$,
with the anisotropy parameterized by $v_3=\frac{\sqrt3 a\gamma_3}{2\hbar}$.
As a result, the Fermi line in bilayer graphene may undergo a topological (Lifshitz) transition\cite{Abrikosov}: from singly connected at high carrier density,
$n_{e(h)}>n_L$ to four separate pockets in momentum space, for $n_{e(h)}<n_L$.
It is common for two-dimensional electron systems\cite{book} that electron-electron (e-e) repulsion
renormalizes the single-particle dispersion of carriers.
In monolayer graphene, where electrons have a characteristic Dirac spectrum, this
leads\cite{singlelayer} to an increase of the Dirac velocity, from $v=\frac{\sqrt3 a\gamma_0}{2\hbar}$ to
$v(p)=v\left(1+\frac{\alpha}{4}\ln(\Lambda/p)\right)$ with $\alpha=e^2/\epsilon_s v$.
In this Communication we investigate the effect of the e-e interaction on the split-band gap $\gamma_1$, the effective mass $m$,
and the dispersion anisotropy parameter $v_3$.
We show for the bilayer that in the Hartree-Fock theory the e-e repulsion increases $\gamma_1$ and $v_3$, but renormalizes
the effective mass\cite{Borghi} and the Lifshitz transition density downwards.
Also, we find that, in contrast to the monolayer, these corrections to the electronic spectrum do not contain infrared divergencies,
and that the renormalization of $n_L$ is weak, so that the single-particle tight-binding model gives a realistic estimation\cite{footfoot} for the
Lifshitz transition density in bilayers.

The theory presented below is based on the four-band model\cite{McCann} describing electrons in bilayer graphene in the vicinity of the
valley centers corresponding to the $K$ ($\xi=1$) and $K'$ ($\xi=-1$) first Brillouin zone corners [see Fig.~\ref{intro}(b)].
The $4\times 4$ tight-binding Hamiltonian,
\begin{equation}
\label{hamilton}
\hat H=\xi\begin{pmatrix}
0 &  v_3\pi & 0 &  v\overline\pi \\
 v_3\overline\pi & 0 &  v\pi & 0 \\
0 &  v\overline\pi & 0 & \xi\gamma_1 \\
 v\pi & 0 & \xi\gamma_1 & 0
\end{pmatrix},\quad
\begin{array}{ll}
\pi&=p_x+ip_y,\\
\overline\pi&=p_x-ip_y,
\end{array}
\end{equation}
which will be the subject of renormalization by the e-e interaction, is written in the basis of
sublattice Bloch states
$[\psi_A,\psi_{\widetilde B},\psi_{\widetilde A},\psi_B]$ in valley $K$ and $[\psi_{\widetilde B},\psi_A,\psi_B,\psi_{\widetilde A}]$ in valley $K'$.
Notice that $\hat H$ has a natural ultraviolet momentum cutoff about $\Lambda\sim1/a$.
This Hamiltonian determines four bands\cite{McCann}, $\pm\epsilon_{1,2}(p)$.
Two split-off bands $\pm\epsilon_2$ start at energies $\pm\gamma_1$.
For small values of momentum, $vp\ll\gamma_1/4$, the two low-energy bands $\pm\epsilon_1$ that touch each other at $K/K'$ can be attributed
to a $2\times2$ effective Hamiltonian\cite{McCann}
\begin{align}
\label{lowenergy}
\hat H'\approx&-\frac{1}{2m}\begin{pmatrix}
0 & \overline\pi^2 \\ \pi^2 & 0
\end{pmatrix}+
\xi v_3\begin{pmatrix}
0 & \pi \\ \overline\pi & 0
\end{pmatrix},
\end{align}
where $m=\frac{\gamma_1}{2v^2}$, and
$\hat H'$ acts on $[\psi_A,\psi_{\widetilde B}]$ in valley $K$ and $[\psi_{\widetilde B},\psi_A]$ in valley $K'$.
The second term in $\hat H'$ causes a triangular distortion of the electronic dispersion illustrated in Fig.~\ref{intro}(b),
where the Lifshitz transition of the electron Fermi line topology is explained,
as a singly-connected line $\epsilon_1(p)=E$ splits into four disconnected pockets at $E<\epsilon_L=\frac{\gamma_1 v_3^2}{2v^2}$, which
would occur at a critical density\cite{footfoot} ($\hbar$ restored)
\begin{equation}
\label{critical}
n_L=\frac{\gamma_1^2}{2\pi\hbar^2 v^2}\left(\frac{v_3}{v}\right)^2.
\end{equation}

In the Hartree-Fock approximation, the change in the single-particle Hamiltonian $\widetilde H=\hat H+\hat\Sigma$ for electrons
can be described using the self-energy diagram
\begin{equation}
\label{sigmadiag}
\hat\Sigma\;=\;\raisebox{-0.5cm}{\includegraphics[width=2cm, keepaspectratio]{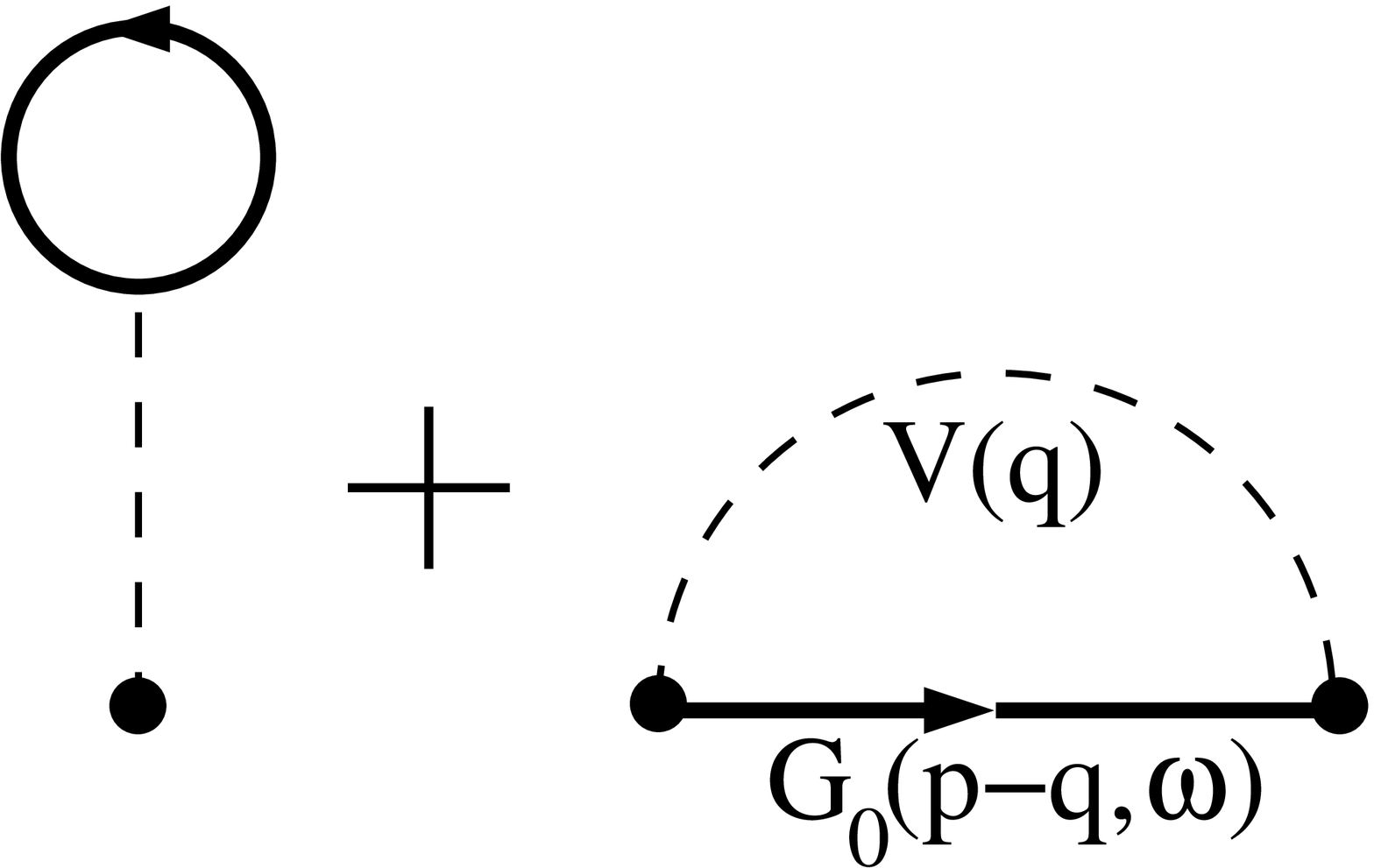}}
\end{equation}
where the solid line denotes the causal Green's function of electrons in the bilayer and the dashed line stands for the interaction
$\widetilde V(q)=2\pi e^2/\epsilon_s q\chi(q)$, where $\epsilon_s$ is the dielectric constant of the environment
($\epsilon_s=\frac{1+4.5}{2}=2.75$ on a SiO$_2$ substrate).
The first (Hartree) diagram in Eq.~(\ref{sigmadiag}), together with the positive background charge cancels the momentum-independent energy shift
determined by the second (Fock) term.
Below, the contribution from the Fock term is estimated both for the bare Coulomb interaction, $V(q)=2\pi e^2/\epsilon_s q$ and for the screened Coulomb interaction,
where the dielectric function $\chi(q)$ is evaluated in the random phase approximation (RPA).
The self-energy $\hat\Sigma$ is evaluated for bilayer graphene at zero doping.
Since the Lifshitz transition occurs at very low doping level, using the renormalized value of the bilayer parameters
that result from this approximation is justified as long as the critical Fermi energy of this transition remains small in comparison to the energy scale $\gamma_1$.

In the $4\times4$ representation the self-energy is a matrix,
\begin{widetext}
\begin{align}
\hat\Sigma(\xi,\mathbf p)=&\int \frac{d\mathbf q}{(2\pi)^2}\frac{\widetilde V(q)}{2(\widetilde\epsilon_1+\widetilde\epsilon_2)}
\left[
\xi v\left(1+\frac{v^2|\mathbf q+\mathbf p|^2}{\widetilde\epsilon_1\widetilde\epsilon_2} \right)
\begin{pmatrix}0 & 0 & 0 & \overline\kappa+\overline\pi \\
0 & 0 & \kappa+\pi & 0 \\
0 & \overline\kappa+\overline\pi & 0 & 0 \\
\kappa+\pi & 0 & 0 & 0
\end{pmatrix}
\right.
+\gamma_1\left(1+\frac{v_3^2|\mathbf q+\mathbf p|^2}{\widetilde\epsilon_1\widetilde\epsilon_2}\right)
\begin{pmatrix}
0 & 0 & 0 & 0\\
0 & 0 & 0 & 0\\
0 & 0 & 0 & 1\\
0 & 0 & 1 & 0
\end{pmatrix}\notag
\\
&-\frac{\gamma_1}{|\mathbf q+\mathbf p|^2}
\begin{pmatrix}0 & (\overline\kappa+\overline\pi)^2 & 0 & 0\\
(\kappa+\pi)^2 & 0 & 0 & 0 \\
0 & 0 & 0 & 0 \\
0 & 0 & 0 & 0
\end{pmatrix}
\left.+\xi v_3\left(1-\frac{\gamma_1^2}{\widetilde\epsilon_1\widetilde\epsilon_2}\right)
\begin{pmatrix}
0 & \overline\kappa+\overline\pi & 0 & 0\\
\kappa+\pi & 0 & 0 & 0\\
0 & 0 & 0 & 0\\
0 & 0 & 0 & 0
\end{pmatrix}+\dots\right].
\label{bigsigma}
\end{align}
Here, we omitted two unimportant terms, and define $\kappa=q_x+iq_y$, $\pi=p_x+ip_y=pe^{i\phi}$,
and $\widetilde\epsilon_{1,2}\equiv\epsilon_{1,2}(\mathbf q+\mathbf p)$ with
\begin{equation}
\label{full}
\epsilon_\alpha^2(\xi,\mathbf p)=\frac{\gamma_1^2}{2}+\left(v^2+\frac{v_3^2}{2}\right) p^2+(-1)^\alpha\sqrt{\frac{1}{4}\left(\gamma_1^2-v_3^2p^2\right)^2+v^2p^2\left(\gamma_1^2+v_3^2p^2\right)+2\xi\gamma_1v_3v^2p^3\cos3\phi}.
\end{equation}
\end{widetext}
The self-energy $\hat\Sigma$ in Eq.~(\ref{bigsigma}) is $\omega$-independent, as usual in the Hartree-Fock approximation.
Schematically, the result of integration over $\mathbf q$ in Eq.~(\ref{bigsigma}) can be represented as
\begin{equation}
\label{sigma}
\hat\Sigma(\xi,\mathbf p)=e^2\begin{pmatrix}
0 & \overline B +Z& 0 & \overline A \\
B+\overline Z & 0 & A& 0 \\
0 & \overline A  & 0 & C \\
A & 0 & C & 0
\end{pmatrix},
\end{equation}
where $A=\xi A_1\pi +O(p^2)$, $C=\frac{\gamma_1}{ v}C_0 + O(p^2)$, $B=\frac{\xi v_3}{v}B_1\overline\pi + \frac{ v}{\gamma_1}B_2\pi^2+O(p^3)$, and $Z=\frac{\xi v_3}{v}M_1\pi+ O(p^3)$.
\begin{table*}[htb]
\begin{center}
\begin{tabular}{c|c|c|c}
\hline\hline
 & bare Coulomb (any $\epsilon_s$) & screened suspended ($\epsilon_s=1$) & screened on SiO$_2$ ($\epsilon_s=2.75$) \\
\hline
$A_1$ & $(0.58\div0.66) + \frac{1}{4}\ln\left(v\Lambda/\gamma_1\right)$ &
$(0.03\div0.04) + 0.032\ln\left(v\Lambda/\gamma_1\right)$ &
$(0.076\div0.098) + 0.071\ln\left(v\Lambda/\gamma_1\right)$\\
$B_1$ & -0.063 & $-0.0013$ & $-0.0034$ \\
$B_2$ & 0.38   & $0.016$ & $0.039$ \\
$C_0$ & $(0\div0.36)+\frac{1}{4}\ln\left(v\Lambda/\gamma_1\right)$ &
$(-0.025\div-0.14)+0.032\ln\left(v\Lambda/\gamma_1\right)$ &
$(-0.047\div-0.021)+0.071\ln\left(v\Lambda/\gamma_1\right)$\\
$M_1$ & $(0.7\div1.1)+\frac{1}{8}\ln\left(v\Lambda/\gamma_1\right)$ &
$(0.036\div0.082) + 0.032\ln\left(v\Lambda/\gamma_1\right)$ &
$(0.09\div0.19) + 0.071\ln\left(v\Lambda/\gamma_1\right)$\\
\hline\hline
\end{tabular}
\end{center}
\caption{\label{renormsigma}
The physical parameters of the self-energy $\hat\Sigma(\xi,\mathbf p)$ (Eq.~(\ref{sigma})), for the bare and screened Coulomb interaction.
Where two limits are given, they correspond to integrating the finite (not logarithmically divergent) part up to $q=\frac{\gamma_1}{v}$
and to $q=\infty$, respectively.
}
\end{table*}
Here we expanded $\hat\Sigma$ in powers of $p$, in order to keep only the terms relevant for the  renormalization
of the bilayer parameters $v,v_3,\gamma_1,m$ and $n_L$.
Term $A$ in Eq.~(\ref{sigma}) takes into account the correction to the intra-layer velocity $v$.
A similar effect is known to occur in monolayers\cite{singlelayer}.
Term $C$ in Eq.~(\ref{sigma}) modifies the inter-layer coupling $\gamma_1$.
The part of terms $B$  and $Z$ that is linear in $\pi,\overline\pi$ yields renormalization of $v_3$.
The quadratic part of term $B$ combines with $A$ and $C$ to determine the renormalization of the effective mass,
\[
\widetilde m^{-1}=\frac{2v^2}{\gamma_1} \left[1+\alpha\left(2A_1-C_0+\frac{1}{2}B_2\right)\right],\quad \alpha=e^2/v\epsilon_s,
\]
which is a result similar to that obtained by Borghi \emph{et al}\cite{Borghi}.
Together, all of these determine the shift of the Lifshitz transition density,
\begin{equation}
\label{critical2}
\widetilde n_L=n_L\left(1+2\alpha\left( C_0+M_1+B_1-2A_1\right) \right).
\end{equation}

The numerical values of the coefficients $A_1$, $B_1$, $B_2$, $C_0$, and $M_1$ in Eq.~(\ref{sigma}) were calculated for suspended graphene flakes
and flakes on SiO$_2$ substrate with and without screening of the Coulomb interaction taken into account\cite{neutral} (Table \ref{renormsigma}).

\begin{figure}[!htbp]
\begin{center}
\includegraphics[width=\columnwidth,keepaspectratio]{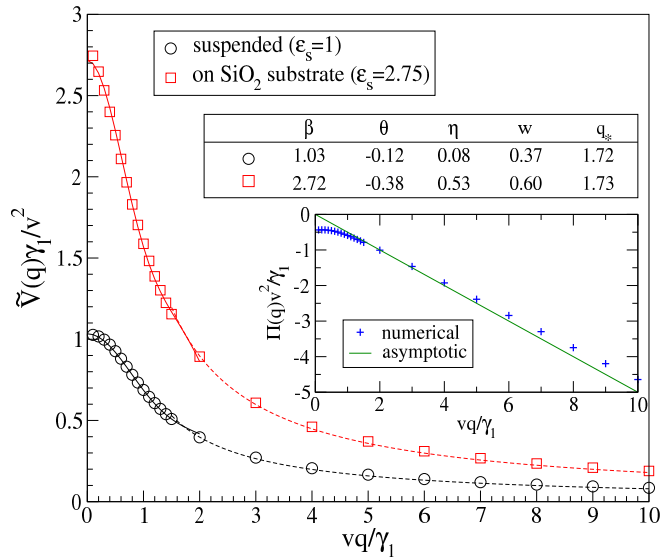}
\end{center}
\caption{\label{polarfig}
(Color online)
The renormalized interaction\cite{neutral} $\widetilde V(q)$ from numerics (symbols), its large momentum asymptotics Eq.~(\ref{asympt}) (dashed line)
and low-momentum asymptotics Eq.~(\ref{approx}) (solid line) for $\epsilon_s=1$ and 2.75.
Inset: the static polarization $\Pi(q)$ from numerics and its asymptotics.
Table: parameters of the small $q$ approximate interaction (Eq.~(\ref{approx})).
}
\end{figure}

Screening of the e-e interaction in graphene is taken into account in the RPA,
by the effective static dielectric function
\begin{align*}
\label{polar}
\chi(\mathbf q)=&1-V(\mathbf q)\Pi(\mathbf q),\\
\Pi(\mathbf q)=&-i\int\frac{d\omega d\mathbf p}{(2\pi)^3}\text{Tr}\left(\hat G_0(\mathbf p+\frac{\mathbf q}{2},\omega)\hat G_0(\mathbf p-\frac{\mathbf q}{2},\omega)\right),
\end{align*}
where $\Pi(\mathbf q)$ is the polarizability of the electron gas.
Fig.~\ref{polarfig} shows the numerical evaluation of $\Pi(q)$, which deviates from the constant value\cite{twoband}
$\approx-0.44\gamma_1/v^2$ obtained earlier using the two-band Hamiltonian $\hat H'$ at $q\sim\gamma_1/2v$,
where the band has a crossover from parabolic to an almost linear behavior.
To obtain analytical asymptotic expressions in the relevant limits first, consider  $q>q_*\gtrsim\gamma_1/v$,
where the dominant contribution comes from $p\gg\gamma_1/v$, and the free electron Green's function reduces to
the free Green's function of two decoupled monolayers.
Then\cite{singlelayer}, $\Pi=-\frac{1}{16}Nq/v$ with $N=8$, and
\begin{equation}
\label{asympt}
\widetilde V(q>q_*)=\frac{2\pi e^2}{(1+\alpha\pi)q}.
\end{equation}
For $q<q_*$, where the electron dispersion is mostly parabolic, screening has the form of Thomas-Fermi
screening with the radius $r_0=0.17v/\gamma_1$ (for $\epsilon_s=1$).
A sufficient simultaneous analytical description of both regimes can be done using an interpolation
formula,
\begin{equation}
\label{approx}
\widetilde V(q<q_*)=(\beta+\theta q^2+\eta q^4)e^{-wq^2},
\end{equation}
where the parameters $\beta,\gamma,\eta,w$ are determined by a numerical fit shown in Fig.~\ref{polarfig},
and the value of $q_*$ is determined from the intersection of $\widetilde V(q<q_*)$ and $\widetilde V(q>q_*)$.

\begin{table}[htb]
\begin{center}
\begin{tabular}{c|c|c|c|c|c}
\hline\hline
           & $\widetilde\alpha$ & $\delta_v$ & $\delta_{v_3}$ & $\delta_{\gamma_1}$ & $\delta_{m^{-1}}$ \\
\hline
I & $\frac{e^2}{\hbar v\epsilon_s}$ & $0.58\div0.66$ & $0.67\div1.05$ & $0\div0.36$     & $1.00\div1.51$\\
II  & 0.28              & $0.07\div0.09$ & $0.08\div0.18$ & $\pm0.05$ & $0.17\div0.20$\\
III & 0.20              & $0.06\div0.08$ & $0.07\div0.15$ & $\pm0.04$ & $0.17\div0.19$\\
\hline\hline
\end{tabular}
\end{center}
\caption{\label{renormall}
Constants for Eq.~(\ref{renormv}) for bare Coulomb interaction (I), screened Coulomb interaction in suspended graphene (II),
and screened Coulomb interaction in graphene on SiO$_2$ substrate (III).\cite{neutral}
Where two limits are given, they correspond to integrating the finite (not logarithmically divergent) part up to $q=\gamma_1/v$ and to $q=\infty$, respectively.
}
\end{table}

Finally, we determine that the bilayer parameters are renormalized by the e-e repulsion as
\begin{align}
\frac{\widetilde v}{v} & = 1+ \frac{\widetilde\alpha Y}{4}   +\delta_v,      \label{renormv}\quad
\frac{\widetilde v_3}{v_3} = 1+ \frac{\widetilde\alpha Y}{8}   +\delta_{v_3},      \\
\frac{\widetilde\gamma_1}{\gamma_1} &= 1+\frac{\widetilde\alpha Y}{4} +\delta_{\gamma_1}, \quad
\frac{\widetilde m^{-1}}{m^{-1}} = 1+ \frac{\widetilde\alpha Y}{4}   +\delta_{m^{-1}}     ,\notag\\
Y&=\ln\left(v\Lambda/\gamma_1\right)\notag,
\end{align}
where the numerical values of all $\delta$'s are listed in Table \ref{renormall}.
This result shows that the intralayer velocity $v$, the interlayer hopping $\gamma_1$,
and the trigonal distortion parameter $v_3$ are all enhanced by the e-e interaction\cite{neutral}.
With bare Coulomb interaction this enhancement would be grossly overestimated if we used the actual
value $\alpha=e^2/\hbar v\approx2.19$ for graphene.

\begin{figure}[htbe]
\begin{center}
\includegraphics[width=\columnwidth, keepaspectratio]{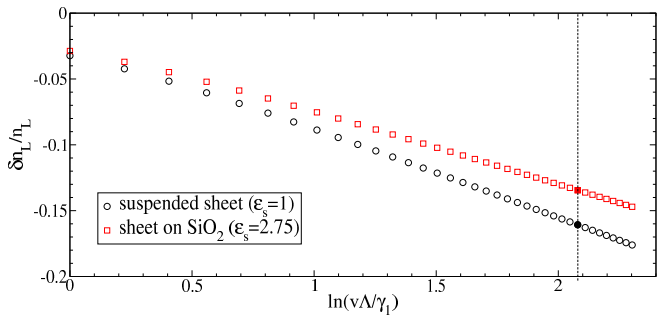}
\end{center}
\caption{\label{reduction}
(Color online)
The relative reduction $\delta n_L/n_L=\widetilde n_L/n_L-1$ of the Lifshitz transition density by the e-e interaction
for a suspended sheet and one on SiO$_2$ substrate.
The vertical line shows the estimation for $\Lambda\approx \gamma_0/v$.
}
\end{figure}

However, after screening is taken into account, we find that one has to usein Eq.~(\ref{renormv}) $\widetilde\alpha\approx 0.28$ for a suspended flake
and $\widetilde\alpha\approx 0.20$ for a flake on SiO$_2$, which gives a much weaker renormalization effect than an estimate using bare Coulomb interaction.
Also, using Eqs.~(\ref{critical2}) and (\ref{renormv}), one may see that in the theory ignoring screening the renormalization of the Lifshitz transition
density $n_L$ would be hugely overestimated, resulting in its disappearance ($n_L\to0$).
However, having taken into account the effect of the reduction of the e-e repulsion by screening, we find
a much smaller shift in the value of $n_L$ from that determined using the tight-binding model for noninteracting electrons\cite{McCann,footfoot}.
The calculated shift in the transition density $n_L$ for the screened Coulomb interaction is shown in Fig.~\ref{reduction}
as a function of momentum cutoff $\Lambda$,
with the vertical line corresponding to $\Lambda\approx\gamma_0/v$.
This determines an approximately 15\% reduction of the Lifshitz transition density for a suspended sheet and about 12\% for a bilayer on a SiO$_2$ substrate.

The result of the above-presented analysis of the bilayer band parameters and the Lifshitz transition density $n_L$ suggests
that their renormalization by the e-e repulsion is relatively weak, due to the screening of the e-e interaction by the electrons themselves.
Thus, we conclude that the Lifshitz transition is not trivially hindered by many-body effects.
The Lifshitz transition can be detected through the singularity of the thermopower\cite{Abrikosov} that develops when a neck
forms from the central pocket to the side pockets of the Fermi line in bilayer graphene (Fig.~\ref{intro}(b)).
Such an observation would require samples of high homogeneity,
but in contrast to bulk metals for which the Lifshitz transition was first discussed,
such an experiment would be possible in graphene since the carrier density in graphene can be directly controlled using external gates.

We thank O.\ Kashuba for useful discussions and help throughout this work.
This work was supported by the Lancaster University-EPSRC Portfolio Partnership.

\newcommand{\PRL}{Phys.\ Rev.\ Lett.}
\newcommand{\PRB}{Phys.\ Rev.\ B}
\newcommand{\PR}{Phys.\ Rev.}
\newcommand{\NPB}{Nucl.\ Phys.\ B}


\begin{thebibliography}{99}
\bibitem{neutral}
Recall that the renormalization by the e-e interaction was calculated for zero doping.
We anticipate these values to be applicable at low doping.
\end{thebibliography}
\end{document}